# Magnetocaloric effect of nanostructured $La_{0.6}Sr_{0.4}CoO_3$

*Fabiana Morales Alvarez[1,2], María Belén Vigna[3], Mariano Quintero[1,2], Diego G. Lamas[4], Joaquín Sacanell[1,2]*

[1]*Departamento de Física de la Materia condensada, Gerencia de Investigación y Aplicaciones, Centro Atómico Constituyentes, CNEA. Av. General Paz 1499, (1650) Villa Maipú, Provincia de Buenos Aires, Argentina.*

[2]*Instituto de Nanociencia y Nanotecnología, Centro Atómico Constituyentes, CNEA-CONICET. Av. General Paz 1499, (1650) Villa Maipú, Provincia de Buenos Aires, Argentina.*

[3]*CONICET / INTEMIN – SEGEMAR, Av. General Paz 5445, (1650) San Martín, Provincia de Buenos Aires, Argentina.*

[4]*Instituto de Tecnologías Emergentes y Ciencias Aplicadas (ITECA), CONICET-UNSAM, Escuela de Ciencia y Tecnología, Laboratorio de Cristalografía Aplicada. Av. 25 de mayo 1169, (1650) San Martín, Provincia de Buenos Aires, Argentina.*

\* Corresponding author: sacanell@tandar.cnea.gov.ar

**Abstract**

In this study, we investigate the magnetic and magnetocaloric properties of nanostructured $La_{0.6}Sr_{0.4}CoO_3$ (LSC) samples synthesized under confinement conditions within porous templates. Using this method, we obtained de-agglomerated nanoparticles, which provide us with the feasibility of applying them in nanoparticle films that can be tailored to intricate geometries. We specifically explored the impact of pore size of the template on key parameters including saturation magnetization ($M_S$), Curie temperature ($T_C$), maximum entropy change ($\Delta S$), and relative cooling power (RCP). Our findings reveal enhancements in those quantities, that are likely to be related with the nanostructure of the samples, indicating the potential of nanostructured LSC as an active material for magnetic refrigeration devices. Our alternative approach of synthesizing magnetocaloric materials under confinement conditions presents an exciting prospect for future research and development in the field.

**Introduction**

The study of the magnetocaloric effect (MCE) in magnetic materials has garnered increasing interest in recent decades, due to a possible application in magnetic refrigeration at room temperature [1,2]. In its simplest form, this effect is defined as the isothermal change in magnetic entropy when a magnetic field is applied to a material. The MCE is mainly characterized by the magnetic entropy change ($\Delta S$) and can be obtained from the M(H) curves measured at different temperatures and then using the following Maxwell relation:

$$\left.\frac{\delta S}{\delta H}\right|_T = \left.\frac{\delta M}{\delta T}\right|_H \tag{1}$$

Or in the integral form,

$$\Delta S(T,H) = \int_0^H \left(\frac{\partial M(H,T)}{\partial T}\right)_H dH' \tag{2}$$

$$\Delta S(T,H) = \int_0^H \left(\frac{[M(T+\Delta T, H')-M(T,H')]}{\Delta T}\right)_H dH' \tag{3}$$

In general, high values of saturation in the magnetization and low coercive fields are an indication of a large change in the magnetic entropy and thus a large MCE, which can be characterized by the relative cooling power (RCP) of the material[1].

In 1997, Pecharsky reported giant MCE at room temperature for Gd and related compounds [3,4] and the interest in the field rapidly increased. However, due to the high cost of Gd related materials, interest has turned to the search of alternative materials such as materials such as $ABO_3$ perovskites, in particular, $La_{1-x}Sr_xMnO_3$ (LSM) and other manganites [5–7] or Ruddelsen-Popper ceramics [8,9], for which a large MCE has been observed close to room temperature.

A family of perovskites with magnetic phenomena similar to that of manganites is the mixed-valence cobalt oxides of the $A_{1-x}A'_xCoO_3$ formula (where A represents a lanthanide and A´ represents a divalent rare earth element). In these compounds, cobalt exhibits various oxidation and spin states, leading to a diverse range of behaviors in terms of their magnetic, electrical, and electrochemical properties. [10–15]

As an example of this richness, in a $Co^{+3}$-$O^{-2}$-$Co^{+3}$ chain, an antiferromagnetic (AFM) interaction is observed, while for $Co^{+3}$-$O^{-2}$-$Co^{+4}$ the interaction is ferromagnetic (FM) and usually coupled with a metallic behavior. A strong competition can occur among those interactions that can be triggered by distortions from the ideal perovskite structure, that arise from the partial substitution of A by A´, giving rise to non-homogeneous magnetic states [16–18].

As a consequence of the above presented scenario, interest has aroused on the study of the magnetic properties and the MCE of Co oxides and related materials [16,18–22]. Those works mostly deal with the fundamentals of MCE in LSC [20,22], and Mn doped LSC [16,21]. In most studies, solid state reaction at high temperatures or sol-gel methods were used to obtain dense ceramic samples with grain sizes in the order of micrometers.

In this study, we investigate the magnetic and magnetocaloric properties of nanostructured LSC. We employed a specific synthesis procedure that enables the straightforward production of deagglomerated nanoparticles. Our research is primarily concentrated on two main aspects: the examination of nanostructured materials, which are relatively underrepresented in the MCE literature for LSC, and the assessment of how the sample's morphology influences the effect. We followed a synthesis procedure that enabled us to obtain deagglomerated nanoparticles, providing the possibility of applying them in nanoparticle films that can be tailored to complex geometries [23].

**Experimental**

$La_{0.6}Sr_{0.4}CoO_3$ nanotubes were synthesized by a pore-wetting technique starting from high-purity $La(NO_3)_3$, $Sr(NO_3)_2$ and $Co(NO_3)_2$ as precursors (Merck, 99.99%). These chemicals were diluted in distilled water and mixed in stoichiometric proportions at a 1 molar concentration in an acidic pH environment. We employed commercial porous polymeric membranes (Isopore™ by Millipore) with two distinct average pore diameters, namely, 200 and 800 nm. These membranes were filled using a syringe system. The resulting samples were labeled as LSC-200 and LSC-800, respectively. The solution within the membranes underwent partial dehydration and denitration using a microwave oven. Subsequently, the polymeric membranes were removed, and the product underwent calcination through thermal treatments at 1000°C. This synthesis procedure is the same as the one outlined in reference [23], which was developed to produce submicrometric diameter tubes with walls composed of nanoparticles. This method represents a straightforward approach to obtaining nanoparticles with significant exposed surface areas [13,14,23,24]. The relative fractions of La and Sr were determined to be 0.583(3) and 0.417(4), respectively, through Rutherford Backscattering analysis.

The structural analysis of the samples was conducted using X-ray powder diffraction (XPD) at the D10B-XPD beamline of the UVX source at the Brazilian Synchrotron Light Laboratory (LNLS) in Campinas, Brazil. A high-intensity (low-resolution) configuration was employed, with a wavelength of 1.7708 Å (corresponding to an incident beam energy of approximately 7000 eV). Rietveld refinements were carried out using the *FullProf Suite* code [25], assuming that the samples exhibited a rhombohedral phase, with the space group $R\bar{3}c$, consisting of ($La^{3+}$, $Sr^{2+}$) cations, $Co^{3+}$ cation and $O^{2-}$ anions occupying the 6a, 6b and 18e positions, respectively. The peak shape was adjusted with a pseudo-Voigt function and the background was fitted by a six-parameter polynomial function in $(2\theta)^n, n = 0 - 5$. Isotropic atomic temperature parameters were applied, and equal thermal parameters were assumed for La and Sr atoms at the A site. No additional constraints or restraints were imposed, maintaining equal values for La and Sr atoms at the A site.

The crystallite size of all the solid solutions was determined using the Scherrer equation, with emphasis on the first intense peak at a low angle to minimize potential microstrain effects. Morphological analysis was conducted through scanning electron microscopy (SEM) employing a Nova™ NanoSEM 230 (Laboratorio de Microscopía Electrónica, Gerencia de Materiales, Centro Atómico Bariloche, CNEA, Argentina). Magnetization measurements were carried out using a Vibrating Sample Magnetometer (VSM) Versalab™ by Quantum Design (Laboratorio de Propiedades Eléctricas y Magnéticas de Óxidos Multifuncionales, Centro Atómico Constituyentes, CNEA, Argentina) to investigate the magnetic and magnetocaloric properties. The samples were cooled to the lowest measurement temperature, ranging from 400 K to 50 K, with the application of a specific magnetic field. Zero-field-cooled-warming (ZFCW) data were recorded during the warming phase. Subsequently, the samples were cooled down to the desired temperature, and field-cooled-cooling (FCC) data were simultaneously collected while sweeping the temperature. Finally, field-cooled-warming (FCW) data were recorded during the warming process. Magnetization was also measured for both full-cycle and first-quadrant variations of the magnetic field.

**Results**

Figure 1 present the XPD data of samples treated at 1000°C. By matching these diffractograms with the entries available in the PDF-2 database of the International Center for Diffraction Data (ICDD), it was determined that they display mostly a rhombohedral perovskite-type crystal structure belonging to the $R\bar{3}c$ space group. However, by means of Rietveld analysis of these diffractograms, it was possible to establish the presence of a small fraction of cubic phase in both samples, as expected considering their small average crystallite sizes [26]. By this analysis, it was possible to determine the relative content of both phases and their respective lattice parameters. The results are presented in Table I together with the average crystallite size, determined by the Scherrer equation, and the standard Rietveld agreement factors.

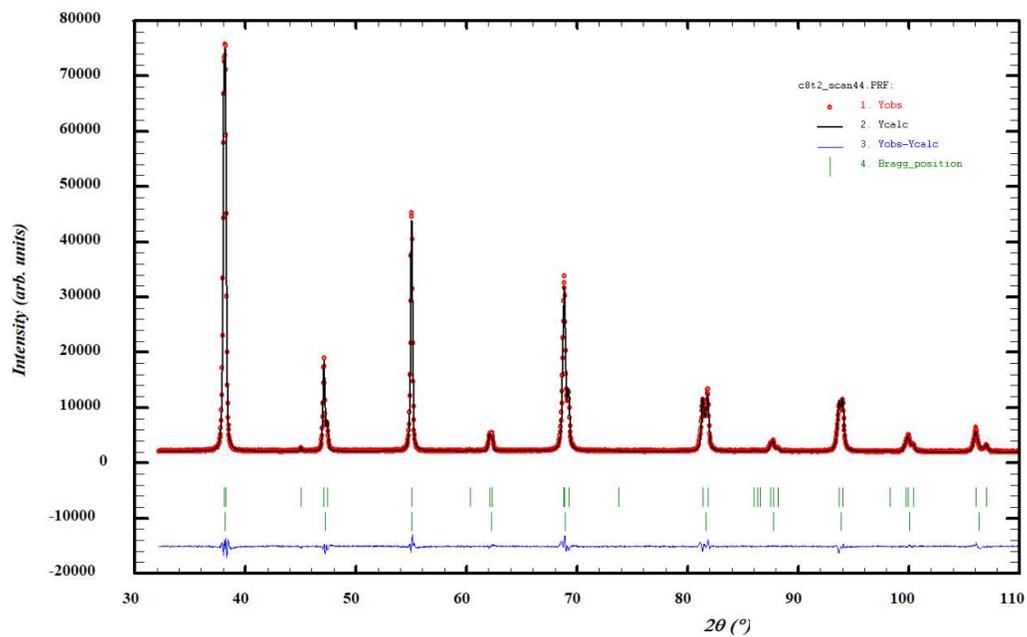

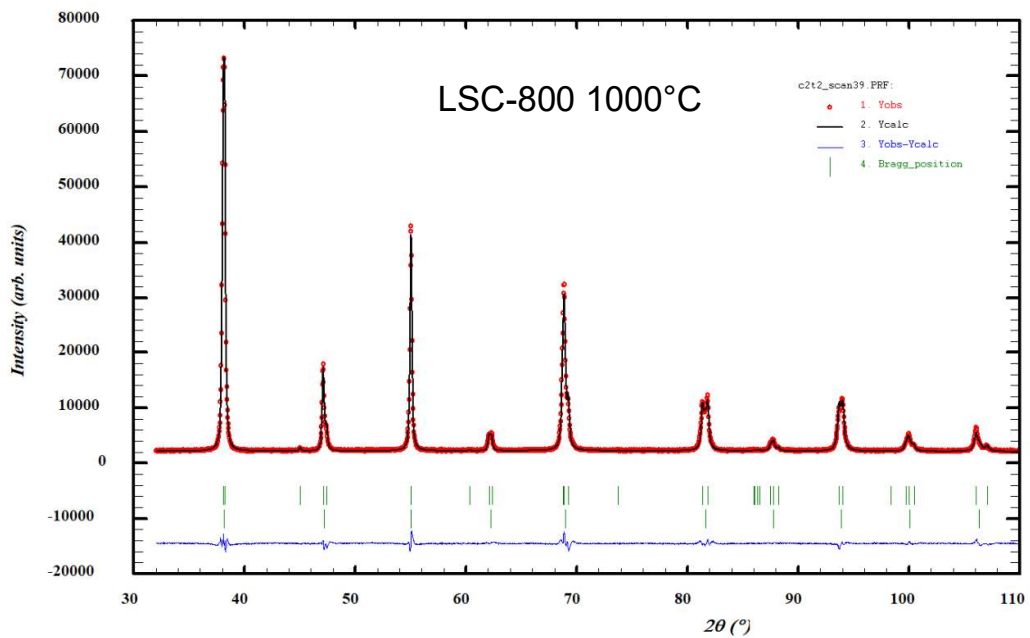

Figure 1. XPD patterns of LSC-200 and LSC-800 samples treated at 1000°C and the calculated patterns determined by Rietveld refinements.

Table I. Results for Rietveld refinements of XPD data for LSC-200 and LSC-800 samples: relative content of the crystalline phases, lattice parameters, average crystallite size and standard Rietveld agreement factors.

| Sample | Relative content (%) | | Lattice parameters (Å) | | Crystallite size (nm) | Rietveld agreement factors |
|---|---|---|---|---|---|---|
| | Cubic | Rhombohedral | Cubic | Rhombohedral | | |
| LSC-200 1000°C | 15.5 (4) | 84.5 (8) | $a = 3.8317\,(1)$ | $a = 5.4327\,(1)$ $c = 13.2101\,(3)$ | 53 (2) | $R_p = 8.17$ ; $R_{wp} = 7.29$ $R_{exp} = 3.50$ ; $\chi^2 = 4.34$ |
| LSC-800 1000°C | 9.6 (4) | 90.4 (9) | $a = 3.8317\,(1)$ | $a = 5.4329\,(1)$ $c = 13.2153\,(3)$ | 56 (2) | $R_p = 7.95$ ; $R_{wp} = 7.18$ $R_{exp} = 3.44$ ; $\chi^2 = 4.37$ |

SEM micrographs of the samples are shown in Figure 2.[1] Two morphologies can be clearly distinguished: The sample synthesized with templates of 200 nm nominal pore size present a rod-like structure, while the material synthesized in 800 nm templates present a tubular morphology. This is due to grain growth experienced by the sample during calcination. From the analysis of SEM micrographs, we determined that the average diameter of the rods was $(135 \pm 13)$ nm. The nanotubes exhibited an average diameter of $(646 \pm 48)$ nm, with a wall thickness of $(73 \pm 17)$ nm. The average size of the particles observed within the rods corresponded closely to their average diameter, while in the case of the nanotubes, it was consistent with the wall thickness. The microstructural analysis presented in our study is the result of an examination of approximately 50 particles for each sample.

---

[1] Technical details, such as Voltage, Magnification, Spot Size, etc, can be viewed in the High Resolution version of the figures, available online.

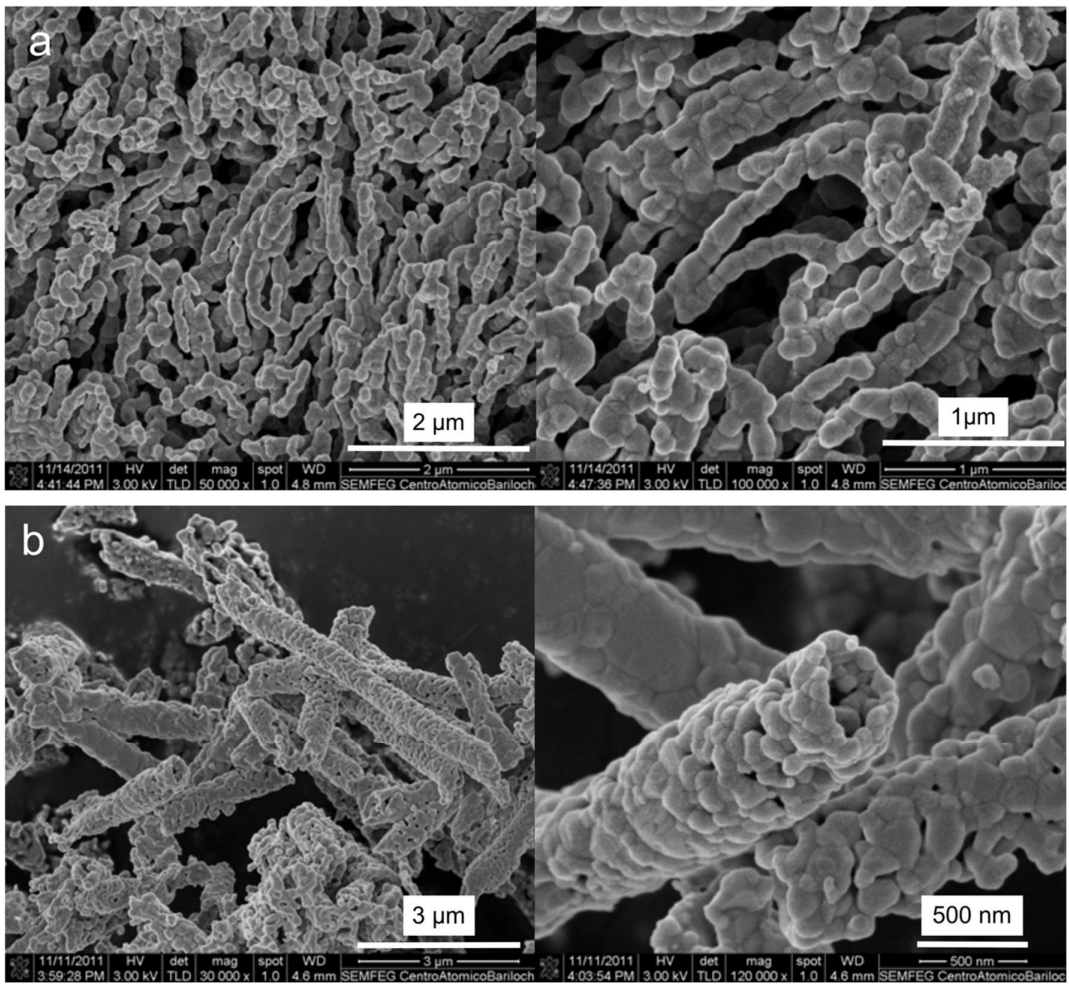

Figure 2. SEM image for samples synthesized in templates of a) LSC-200 nm, b) LSC-800 nm.

Figure 3 depicts magnetization versus magnetic field (M vs H) measurements for both samples at three different temperatures: 50 K (a), 240 K (b), and 300 K (c), with the magnetic field varying from +30 kOe to -30 kOe. At low temperatures, hysteresis loops are observed, while at 300 K, a paramagnetic-like behavior is evident. The samples display open hysteresis cycles at 50K, well below the range in which the most pronounced alteration in magnetic entropy becomes apparent. This significant gap underlines that the observed hysteresis losses occur well beyond the temperature span relevant to the material's potential applications. The saturation magnetization ($M_S$) values are presented in Table II, indicating that the $M_S$ values for LSC-200 are slightly smaller than those for LSC-800. A similar trend is observed for remanence ($M_R$) and coercive field ($H_C$). These values play a significant role in the magnetocaloric effect (MCE): higher $M_S$ and $M_R$ values suggest a larger MCE, and the moderate $H_C$ is important as it indicates a reversible M vs H relationship. At 240 K (Fig. 3b), a decrease in $M_S$ is observed for all samples compared to the values at 50 K. In Figure 3c, a linear dependence is evident, signifying a magnetic transition from ferromagnetic to paramagnetic behavior.

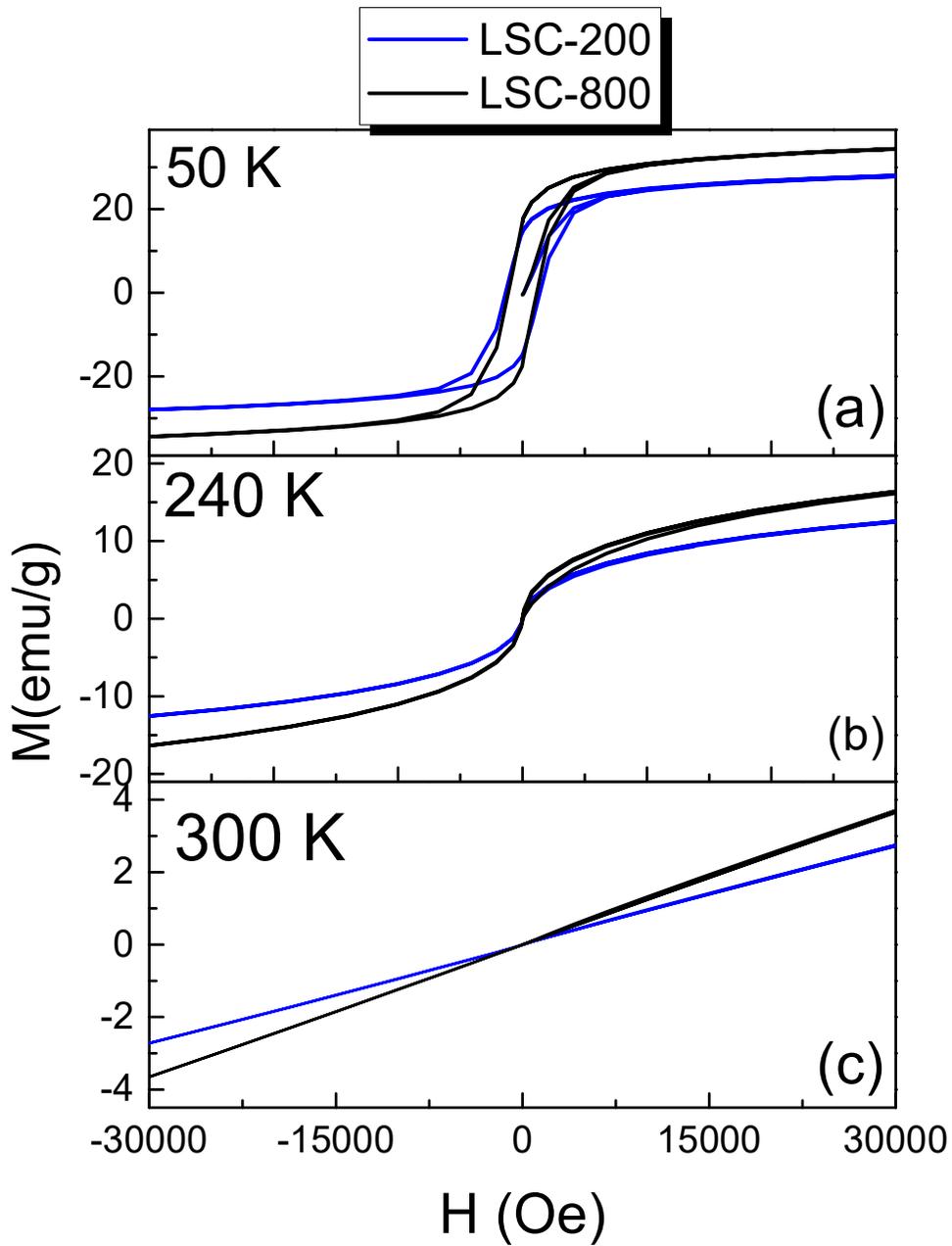

Figure 3: Magnetization vs Magnetic Field for LSC samples synthesized with templates of 200 nm and 800 nm, at (a) 50 K, (b) 240 K and (c) 300 K.

Figure 4 shows magnetization as a function of the temperature in modes zero field cooling (ZFC), field cooled cooling (FCC) and field cooled warming (FCW), with applied magnetic fields of 100

Oe and 1000 Oe. A single transition from a paramagnetic state to a ferromagnetic state occurs on cooling and a blocking temperature ($T_B$) can be observed for all samples, indicating a superparamagnetic behavior. Transition and blocking temperatures are presented in Table II. The existence of a blocked behavior is a clear indication that we are dealing with nanostructured samples.

The values of blocking temperatures corresponding to LSC-200 and LSC-800 are very similar. The Curie transition temperatures ($\theta$) were obtained by fitting the inverse of the high temperature range, assuming a Curie-Weiss (CW) law $\chi = \frac{C}{(T-\theta)}$, where $C$ is the Curie constant. [1,15] The obtained values show a measurable increase from LSC-200 to LSC-800, as shown in Table II.

The experimental effective PM moment is calculated using the expression $\mu_{eff}^{exp}(\mu_B) = \left(\frac{3Ck_B}{N_a \mu_B^2}\right)^{\frac{1}{2}} = \sqrt{8C}$, where $N_a$ is the Avogadro number, $k_B$ is the Boltzmann constant and $C$ can be obtained from the fitted curve. [27,28]

The values for $\mu_{eff}$ are both around 3 $\mu_B$ with no significant differences. The obtained values are in accordance with [13]. Thus, the disparity observed in the Ms values is likely to be related to the different relative content of cubic and rhombohedral phases in the samples.

Table II. Summary of the magnetic parameters of the LSC samples. $M_S$, $M_R$ and $H_C$, measured at 50 K. The Curie and the blocking temperatures have been measured at 1000 Oe.

|  | $M_s\,(emu/g)$ | $M_R\,(emu/g)$ | $H_C\,(Oe)$ | $\theta\,(K)$ | $T_B\,(K)$ | $\mu_{eff}^{exp}(\mu_B)$ |
|---|---|---|---|---|---|---|
| LSC-200 | 27±1 | 14.46±0,03 | 1396 ± 44 | 254.0±0,5 | 205±1 | 3.05±0.02 |
| LSC-800 | 31.8 ±0,7 | 16.349±0,002 | 1204.8±0.1 | 269.2±0,5 | 207±1 | 3.02±0.02 |

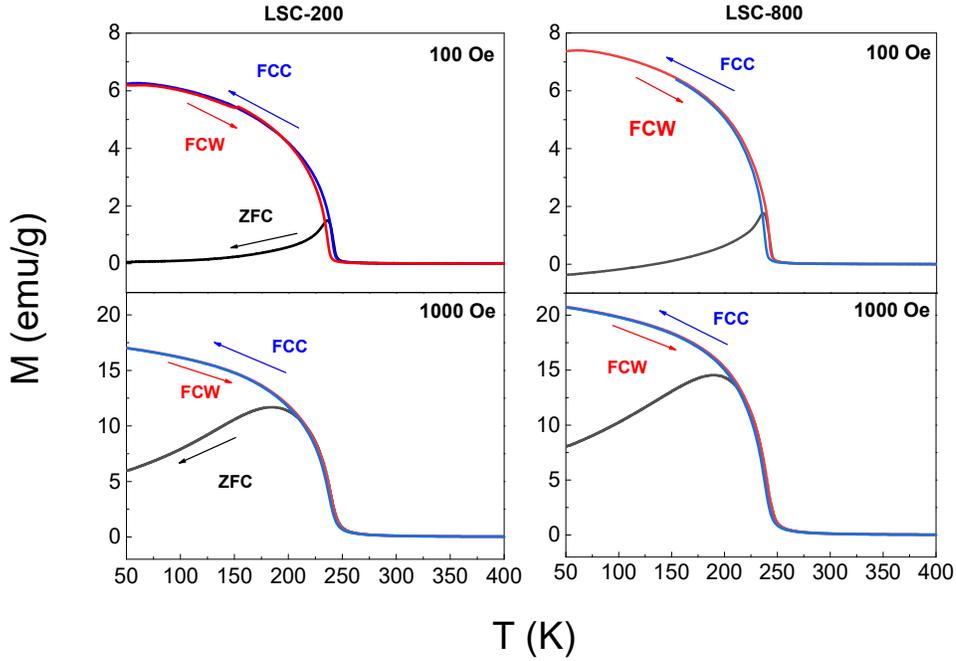

Figure 4: Magnetization vs temperature of LSC samples measured at 100 Oe and 1000 Oe.

Table III. Magnetocaloric parameters. Magnetic entropy and RCP are presented for H =30 kOe.

| Sample | $T_{peak}$ (K) | $-\Delta S_M^{Max}\ \frac{J}{kg/K}$ | $RCP\ (\frac{J}{kg})$ |
|---|---|---|---|
| LSC-200 | 240±5 | 0,77±0,001 | 35±4 |
| LSC-800 | 240±5 | 1,13±0,001 | 58±5 |

In Figure 5, we depict the relationship between $M^2\ vs\ H/M$, commonly referred to as an Arrott plot. Following the criteria outlined in references [29,30], this plot allows us to discern the nature of the magnetic transition. Curves characterized by negative slopes indicate a first-order magnetic transition, while those with positive slopes suggest a second-order magnetic transition.

In our specific study, the slopes derived from the data for the four samples consistently exhibit a positive trend. This observation strongly implies a second-order magnetic transition occurring from the paramagnetic to the ferromagnetic state. Remarkably, we can further determine the Curie temperature ($\theta$) through extrapolation of these curves. This is accomplished by identifying the point at which the curve intersects the zero line, providing valuable insight into the critical temperature of the magnetic phase transition.

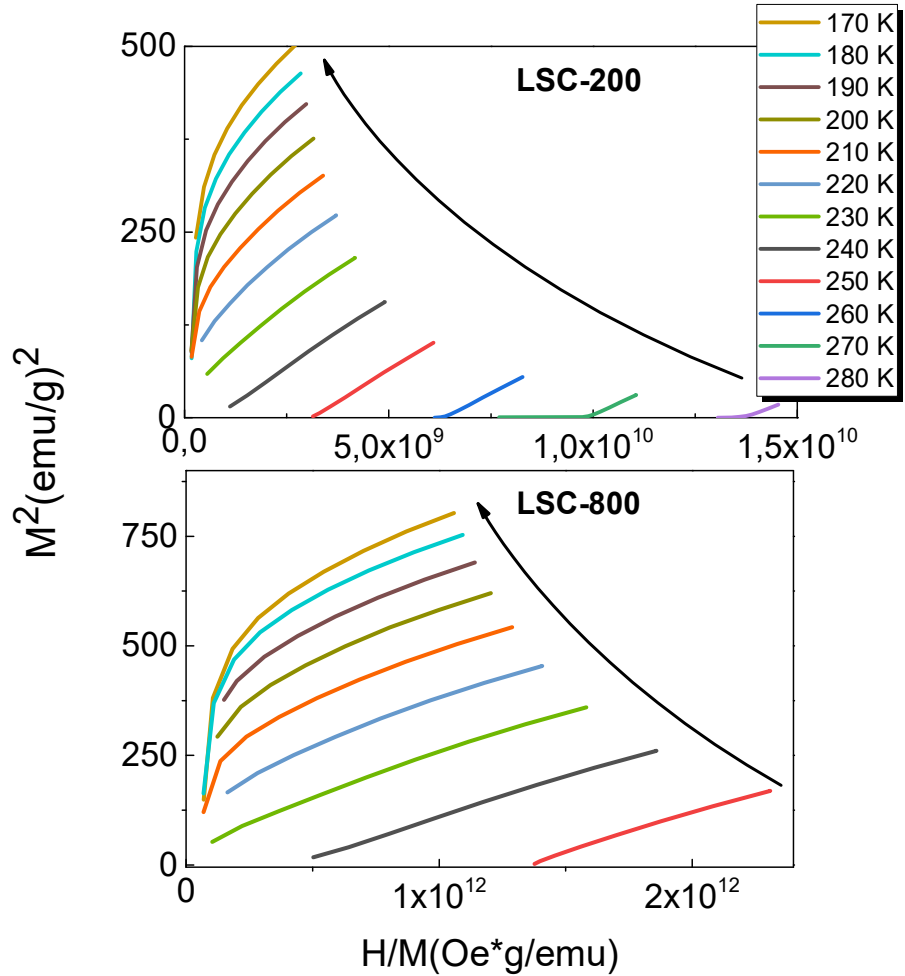

Figure 5. Arrott plot for LSC-200 and LSC-800 samples

Using Maxwell's relations (Eq. 3), we derived entropy values by integrating the M(H) data obtained under various applied magnetic fields (10000 Oe, 20000 Oe, and 30000 Oe). The results are illustrated in Figure 5.

For each sample, we observed a distinct maximum in entropy at a specific temperature denoted as $T_{peak}$ for each sample. The values of $T_{peak}$ were determined as the temperatures at which the maximum entropy change ($\Delta S_{Max}$) was achieved, and they are presented in table III.

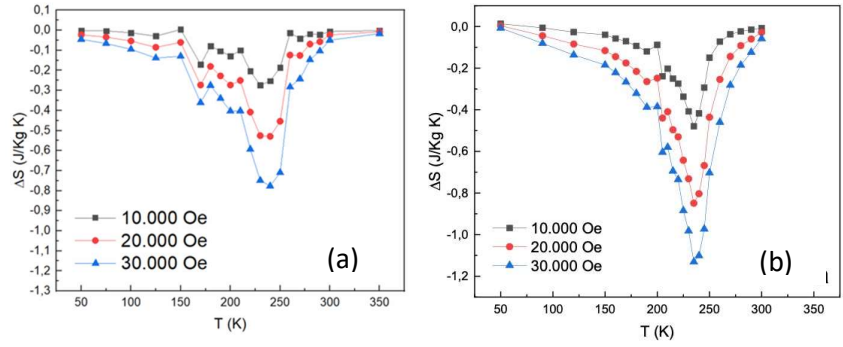

Figure 5. Temperature dependent $\Delta S_M$ curves for (a) LSC-200 and (b) LSC-800 samples with magnetic field of 10000, 20000 and 30000 Oe.

With the results for the entropy value $\Delta S$ we can evaluate the efficiency of a magnetocaloric material which is the relative cooling power (RCP), which is a measure of magnetic cooling efficiency. This refers to the amount of heat that can be transferred by the material from a cold source to a hot source, considering an ideal refrigeration cycle and is given by:[31,32]

$$RCP = \Delta S_{Max} \times \delta F_{FWFM} \quad (4)$$

Where $\Delta S_{Max}$ is the maximum value of entropy at $T_C$ and $\delta F_{FWFM}$ is the full width of the of the entropy curve peak. The RCPs calculated for all samples are presented in Figure 6, with notably higher values observed for the LSC-800 samples in comparison to the LSC-200 samples. While it cannot be definitively confirmed, a plausible explanation for this difference is likely associated with the varying relative fractions of cubic and rhombohedral phases present in the samples. In this context, it appears that the rhombohedral phase exhibits the highest magnetocaloric effect (MCE). Nevertheless, further research is required to substantiate this observation.

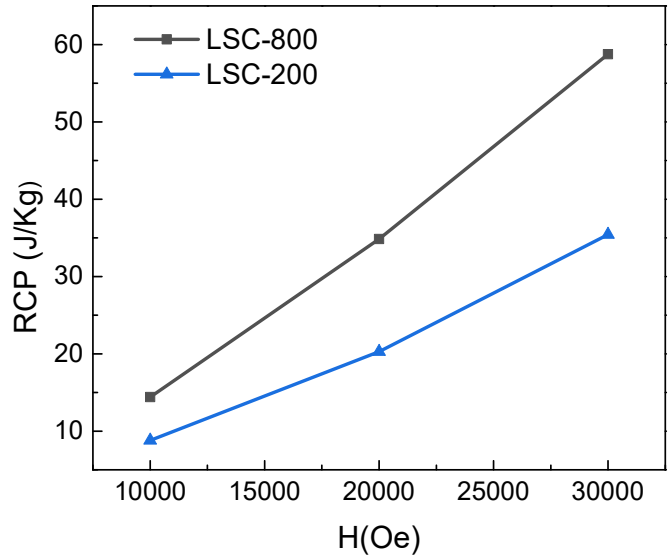

Fig. 6. Relative cooling power for all samples vs. magnetic field.

Both the M vs H and M vs T measurements indicate the presence of magnetic correlations for T < 240 K. The discrepancy between the FC and ZFC measurements suggests a superparamagnetic behavior with elevated saturation magnetization values. Furthermore, the relatively low coercive field observed in both samples indicates easy cycles of magnetization and demagnetization. It is essential to emphasize that both of these characteristics are advantageous for the magnetocaloric effect in the material.

Moreover, it is noteworthy that both the blocking and Curie temperatures decrease when subject to reduced heat treatment (not shown) or when employing smaller templates, signifying that samples with smaller particles will exhibit lower transition temperatures.

As the applied magnetic field increases, the value for the peak change in entropy also rises, and $T_{peak}$ exhibits a slight shift towards higher temperatures. The largest value is the corresponding to sample LSC-800, which reaches a value of 1.13 $Jkg^{-1}K^{-1}$ at around 240 K. Biswal et al. [16] and Long et al. [17] report an entropy change of 2.6 $Jkg^{-1}K^{-1}$ for the compound under a magnetic field of 5T, while Saadaoui et al. [20] report 1.18 $Jkg^{-1}K^{-1}$. The maximum entropy change obtained for the LSC-800 sample (Table III) is similar to those reported for the compound. Our samples are composed of non-agglomerated nanoparticles, whereas the mentioned references involve materials with a high degree of agglomeration. The reported values remain considerably lower than those reported for Gd-based compounds [31], where the primary drawback lies in their cost.

**Conclusions**

In summary, we have studied the magnetic and magnetocaloric behavior in $La_{0.6}Sr_{0.4}CoO_3$ perovskites synthesized in porous templates (LSC-200 and LSC-800). We obtained an arrange of submicrometric tubes and rods formed by nanoparticles. Samples synthesized in templates with larger pores exhibit an increase in the Curie temperature, together with an increase in the saturation of the magnetization. Furthermore, we observed hysteresis in LSC-800 with a higher remanent magnetization compared to LSC-200.

We found that the maximum values of the entropy change are reached slightly below the Curie temperature of the samples, with LSC-800 exhibiting the most significant effect. Finally, there is a favorable impact of the increase in the used pore size, both on the values of $\Delta S_M^{Max}$ and RCP. The $\Delta S_M^{Max}$ values and RCP are similar and are in the same order as those reported in the literature. Those combined facts suggest that the observed favorable impact is most likely to be due to nanostructuration.

We identified an interesting feature for future research in the possibility of modifying the various magnetic parameters ($M_S$, $M_R$, $H_C$, Curie Temperature, etc.) through the partial substitution of Co with different magnetic ions. As Mn has been already studied, we propose to explore the use of Fe. Further investigation into this topic could provide valuable insights and contribute to the advancement of magnetocaloric materials.

Our study demonstrates the potential of $La_{0.6}Sr_{0.4}CoO_3$ perovskites synthesized in porous templates for magnetic refrigeration applications. The significant values obtained for key magnetic and magnetocaloric properties, coupled with the possibility of manipulating these properties through the substitution of magnetic ions, open up exciting avenues for future research and development in the field. These findings pave the way for the design and optimization of advanced magnetic refrigeration devices, offering energy-efficient and environmentally friendly cooling solutions. Our work contributes to the understanding of confined synthesis techniques and provides valuable insights for the advancement of magnetocaloric materials.

The results indicate that the rhombohedral phase possesses enhanced properties from the perspective of the Magnetocaloric Effect (MCE), as the sample with a higher content of this phase exhibits increased parameters that positively influence the MCE. Thus, another interesting point for further investigation is the modification of the relative content of the cubic and rhombohedral phases, in order to control MCE.

**Acknowledgments**

The present work was partially supported by the Brazilian Synchrotron Light Laboratory (LNLS, Brazil, proposal XPD 11900) and Agencia Nacional de Promoción Científica y Tecnológica (Argentina, projects PICT 2017 N° 858, PICT 2018 N°2397 and 3021, and PICT 2020 SERIE A N° 03759).

**References**

[1] N. Raghu Ram, M. Prakash, U. Naresh, N. Suresh Kumar, T. Sofi Sarmash, T. Subbarao, R. Jeevan Kumar, G. Ranjith Kumar, K. Chandra Babu Naidu, Review on Magnetocaloric Effect and Materials, J. Supercond. Nov. Magn. 31 (2018) 1971–1979. https://doi.org/10.1007/s10948-018-4666-z.

[2] A. Greco, C. Aprea, A. Maiorino, C. Masselli, A review of the state of the art of solid-state caloric cooling processes at room-temperature before 2019, Int. J. Refrig. 106 (2019) 66–88. https://doi.org/10.1016/j.ijrefrig.2019.06.034.

[3] V.K. Pecharsky, K.A. Gschneidner, Effect of alloying on the giant magnetocaloric effect of Gd5(Si2Ge2), J. Magn. Magn. Mater. 167 (1997) L179–L184. https://doi.org/10.1016/S0304-8853(96)00759-7.

[4] S.Y. Dan'kov, A.M. Tishin, V.K. Pecharsky, K.A. Gschneidner, Magnetic phase transitions and the magnetothermal properties of gadolinium, Phys. Rev. B. 57 (1998) 3478–3490. https://doi.org/https://doi.org/10.1103/PhysRevB.57.3478.

[5] S. Passanante, L.P. Granja, C. Albornoz, D. Vega, D. Goijman, M.C. Fuertes, Journal of Magnetism and Magnetic Materials Magnetocaloric effect in nanocrystalline manganite bilayer thin films, J. Magn. Magn. Mater. 559 (2022). https://doi.org/10.1016/j.jmmm.2022.169545.

[6] M.H. Phan, S.C. Yu, Review of the magnetocaloric effect in manganite materials, J. Magn. Magn. Mater. 308 (2007) 325–340. https://doi.org/10.1016/j.jmmm.2006.07.025.

[7] D. Goijman, A.G. Leyva, M. Quintero, Measuring magnetocaloric effect in a phase separated system, Mater. Res. Express. 6 (2018) 026106. https://doi.org/10.1088/2053-1591/aaf04a.

[8] A. Kumar, K. Kumari, S. Minji, M.K. Sharma, Z. Zhang, S.H. Huh, B.H. Koo, Excellent cooling power in chemically compressed double layer Ruddlesden-Popper ceramics La1.4-xNdxSr1.6Mn2O7 (0.0 ≤ x ≤ 0.15), Ceram. Int. 48 (2022) 4626–4636. https://doi.org/10.1016/J.CERAMINT.2021.10.249.

[9] A. Kumar, A. Vij, S.H. Huh, J.W. Kim, M.K. Sharma, K. Kumari, N. Yadav, F. Akram, B.H. Koo, Evidence of a moderate refrigerant capacity in cation disordered Ruddlesden-Popper compounds A1.4Sr1.6Mn2O7 (A = La, Pr, Nd) probed with various figures of merit, Curr. Appl. Phys. 49 (2023) 35–44. https://doi.org/10.1016/J.CAP.2023.02.014.

[10] M.A. Señarís-Rodríguez, J.B. Goodenough, Magnetic and Transport Properties of the System La1-xSrxCoO3-δ (0 < x ≤ 0.50), J. Solid State Chem. 118 (1995) 323–336. https://doi.org/10.1006/jssc.1995.1351.

[11] G.H. Jonker, J.H. Van Santen, Magnetic compounds wtth perovskite structure III.

ferromagnetic compounds of cobalt, Physica. 19 (1953) 120–130. https://doi.org/10.1016/S0031-8914(53)80011-X.

[12] V.G. Bhide, D.S. Rajoria, C.N.R. Rao, G.R. Rao, V.G. Jadhao, Itinerant-electron ferromagnetism in La1-xSrxCoO3: A Mossbauer study, Phys. Rev. B. 12 (1975) 2832–2843. https://doi.org/10.1103/PhysRevB.12.2832.

[13] A.E. Mejía Gómez, J. Sacanell, C. Huck-Iriart, C.P. Ramos, A.L. Soldati, S.J.A. Figueroa, M.H. Tabacniks, M.C.A. Fantini, A.F. Craievich, D.G. Lamas, Crystal structure, cobalt and iron speciation and oxygen non-stoichiometry of La0.6Sr0.4Co1-yFeyO3-δ nanorods for IT-SOFC cathodes, J. Alloys Compd. 817 (2020). https://doi.org/10.1016/j.jallcom.2019.153250.

[14] A. Mejía Gómez, J. Sacanell, A.G. Leyva, D.G. Lamas, Performance of La0.6Sr0.4Co1−yFeyO3 (y=0.2, 0.5 and 0.8) nanostructured cathodes for intermediate-temperature solid-oxide fuel cells: Influence of microstructure and composition, Ceram. Int. 42 (2015) 3145–3153. https://doi.org/10.1016/j.ceramint.2015.10.104.

[15] N. Menyuk, P.M. Raccah, K. Dwight, Magnetic Properties of La0.5Sr0.5CoO3, Phys. Rev. 166 (1968) 510–513. https://doi.org/10.1103/PhysRev.166.510.

[16] H. Biswal, T.R. Senapati, A. Haque, J.R. Sahu, Beneficial effect of Mn-substitution on magnetic and magnetocaloric properties of La0.5Sr0.5CoO3 ceramics, Ceram. Int. 46 (2020) 11828–11834. https://doi.org/10.1016/j.ceramint.2020.01.217.

[17] P.T. Long, T. V. Manh, T.A. Ho, V. Dongquoc, P. Zhang, S.C. Yu, Magnetocaloric effect in La1-xSrxCoO3 undergoing a second-order phase transition, Ceram. Int. 44 (2018) 15542–15549. https://doi.org/10.1016/J.CERAMINT.2018.05.216.

[18] L.T.T. Ngan, N.T. Dang, N.X. Phuc, L. V. Bau, N. V. Dang, D.H. Manh, P.H. Nam, L.H. Nguyen, P.T. Phong, Magnetic and transport behaviors of Co substitution in La0.7Sr0.3MnO3 perovskite, J. Alloys Compd. 911 (2022). https://doi.org/10.1016/j.jallcom.2022.164967.

[19] P.T. Long, T. V. Manh, T.A. Ho, V. Dongquoc, P. Zhang, S.C. Yu, Magnetocaloric effect in La1-xSrxCoO3 undergoing a second-order phase transition, Ceram. Int. 44 (2018) 15542–15549. https://doi.org/10.1016/j.ceramint.2018.05.216.

[20] F. Saadaoui, R. M'nassri, H. Omrani, M. Koubaa, N.C. Boudjada, A. Cheikhrouhou, RSC Advances Critical behavior and magnetocaloric study in La 0.6 Sr 0.4 CoO 3 cobaltite prepared by a sol–gel process Article, RSC Adv. 6 (2016) 50968–50977. https://doi.org/10.1039/C6RA08132K.

[21] R. Tetean, I.G. Deac, E. Burzo, A. Bezergheanu, Magnetocaloric and magnetoresistance properties of La2/3Sr1/3Mn1-xCoxO3 compounds, J. Magn. Magn. Mater. 320 (2008) 179–182. https://doi.org/10.1016/j.jmmm.2008.02.100.

[22] R. Li, P. Kumar, R. Mahendiran, Critical behavior in polycrystalline La0.7Sr0.3CoO3

from bulk magnetization study, J. Alloys Compd. 659 (2016) 203–209. https://doi.org/10.1016/j.jallcom.2015.11.060.

[23] J. Sacanell, A.G. Leyva, M.G. Bellino, D.G. Lamas, Nanotubes of rare earth cobalt oxides for cathodes of intermediate-temperature solid oxide fuel cells, J. Power Sources. 195 (2010) 1786–1792. https://doi.org/10.1016/j.jpowsour.2009.10.049.

[24] A.E. Mejía Gómez, D.G. Lamas, A.G. Leyva, J. Sacanell, Nanostructured $La_{0.5}Ba_{0.5}CoO_3$ as cathode for solid oxide fuel cells, Ceram. Int. 45 (2019) 14182–14187. https://doi.org/10.1016/j.ceramint.2019.04.122.

[25] J. Rodríguez-Carvajal, Recent advances in magnetic structure determination by neutron powder diffraction, Physica B: Condensed Matter 192 (1993) 55–69. https://doi.org/10.1016/0921-4526(93)90108-I.

[26] L.M. Acuña, J. Peña-Martínez, D. Marrero-López, R.O. Fuentes, P. Nuñez, D.G. Lamas, Electrochemical performance of nanostructured $La_{0.6}Sr_{0.4}CoO_{3-\delta}$ and $Sm_{0.5}Sr_{0.5}CoO_{3-\delta}$ cathodes for IT-SOFCs, J. Power Sources. 196 (2011) 9276–9283. https://doi.org/10.1016/j.jpowsour.2011.07.067.

[27] D. Kumar, P. Jena, A.K. Singh, Structural, magnetic and dielectric studies on half-doped $Nd_{0.5}Ba_{0.5}CoO_3$ perovskite, J. Magn. Magn. Mater. 516 (2020) 167330. https://doi.org/10.1016/J.JMMM.2020.167330.

[28] H. Biswal, V. Singh, R. Nath, S. Angappane, J.R. Sahu, Magnetic and magnetocaloric properties of $LaCr_{1-x}Mn_xO_3$ (x = 0, 0.05, 0.1), Ceram. Int. 45 (2019) 22731–22736. https://doi.org/10.1016/j.ceramint.2019.07.311.

[29] A. Arrott, Criterion for ferromagnetism from observations of magnetic isotherms, Phys. Rev. 108 (1957) 1394–1396. https://doi.org/10.1103/PhysRev.108.1394.

[30] B.K. Banerjee, On a generalised approach to first and second order magnetic transitions, Phys. Lett. 12 (1964) 16–17. https://doi.org/10.1016/0031-9163(64)91158-8.

[31] A. Gschneidner, V.K. Pecharsky, A.O. Tsokol, Recent developments in magnetocaloric materials, Reports Prog. Phys. 68 (2005) 1479–1539. https://doi.org/10.1088/0034-4885/68/6/R04.

[32] K.A. Gschneidner, V.K. Pecharsky, Magnetocaloric materials, Annu. Rev. Mater. Sci. 30 (2000) 387–429. https://doi.org/10.1146/annurev.matsci.30.1.387.